\documentclass[a4paper,12pt]{aastex}
\usepackage{spr-astr-addons}
\usepackage{url}

\RequirePackage{color}

\newcommand{\emaila}{binayrai21@gmail.com}

\begin{document}

\title{Timing and spectral properties of Be/Xray pulsar 4U 1901+03 during 2019 outburst}
\slugcomment{Not to appear in Nonlearned J., 45.}
\shorttitle{4U 1901+03}
\shortauthors{Binay Rai et al.}

\author{Binay Rai\altaffilmark{1}} \and \author{Bikash Chandra Paul \altaffilmark{1}}
\email{\emaila}

\altaffiltext{1}{Department of Physics \\ North Bengal University \\ Raja rammohanpur, Darjeeling, W. B. -734013\\ E-mail:binayrai21@gmail.com (BR)\\ bcpaul@associates.iucaa.in (BCP)}

\abstract{We have studied the timing and spectral properties of the BeXB 4U 1901+03 during the 2019 outburst using \textit{NuSTAR}, \textit{Swift}, and \textit{NICER} observations. Flares are in all observations and were of tens to hundreds of seconds duration. Pulse profiles were changing significantly with time and the luminosity of the source. An increase in the height of the peak of the pulse profiles was observed with energy. The pulse fraction increases with energy and at the end of the outburst. The variation of the pulse profile with time indicates the transition of the pulsar in different accretion regimes. The absorption like feature at 10 keV shows a positive correlation with the luminosity and along with other spectral parameters this feature was also pulse phase dependent. As the distance to the source is not precisely known hence we cannot confirm this feature to be CSRF and also cannot ignore other possible explanations of the feature. Another absorption like feature about 30 keV was observed in the spectra of the last two \textit{NuSTAR} observations and has line energy of about 30.37$\pm$0.55 and 30.23$\pm$0.62 keV respectively. We have also studied the variation of the line energy, width, and optical depth of this feature with pulse phase. The softening of the spectrum along with the increase in pulse fraction at the end of the outburst and absence of pulsation after 58665.09 MJD suggest that pulsar has entered propeller phase, also abrupt decrease in \textit{Swift}-XRT flux supports the fact.



\section{Introduction}
X-ray binaries (XRB) falls under the broad class of binary stars. One of the component of X-ray binary is compact object  (white dwarf, neutron star and black hole). Depending upon the mass of the companion star, X-ray binaries are classified into High Mass X-ray Binaries (HMXB) and Low Mass X-ray Binaries(LMXB). XRBs with neutron star is further classified into Be X-ray Binaries (BeXB) and Super Giant X-ray Binaries (SGXB) \citep{Reig11}. BeXBs consist of normal Be star along with the neutron star. They are mostly transient in nature and are observable during a bright outburst.

The X-ray source 4U 1901+03 is a BeXB which was first detected in 1970-1971 by Uhuru mission. The source was not observed  for few decades but it was finally observed in 2003 when the source appeared again during a giant outburst \citep{Galloway05}. The \textit{Rossi X-ray Timing Explorer} observations of 2003 outburst revealed the source to be a pulsar with a pulse period of 2.763 s and orbital period of 22.58 days \citep{Galloway05}. \cite{James11} observed X-ray flares, broadening of pulse frequency and qausi-periodic oscillation (QPO) in the source. The flares were 100-300 s long lasting, stronger and more frequently observed during the peak of the outburst. The frequency of QPO is centered around $\sim$0.135 Hz with the r.m.s. value of 18.5$\pm$3.1 per cent. 

The residue of the best fitted spectra of the source showed a significant deviation near 10 keV, this 10 keV feature were observed during 2003 by \cite{Reig16, Galloway05} in the spectra of the pulsar. This feature was found to be dependent on the flux \cite{Reig16} and possibly be a cyclotron line. The pulse profile of a pulsar varies throughout an outburst and indicates different accretion regimes \citep{Basko76,Becker12,Mushtukov15} ,the study of the pulse profiles of 4U 1901+03 predicted that the object passed through different accretion regimes during the outburst \cite{Reig16}. These regimes are defined by a certain value of luminosity called critical luminosity \citep{Basko76,Becker12,Mushtukov15}. The critical luminosity ($L_{crit}$) is define as luminosity above which the radiation pressure is strong enough to stop the accreting matter at a certain distance above the neutron star. The super-critical regime is reached when the luminosity of the pulsar ($L_{X}$) is greater than $L_{crit}$. In this case radiation dominated shock wave is formed which moves up to few kilometers above the neutron star. However for sub-critical regime $L_{X}<L_{crit}$, accreting material are capable of reaching onto the surface of neutron star with heating it. In case of the super-critical regime X-ray photons escape from the side surface of the accretion column perpendicular to the magnetic field lines thus forming fan shaped beam but for sub-critical regime the emission is parallel to the magnetic field which come out as a  pencil beam pattern consisting of pulsed component with simple pulse profile. The pulse profile associated with the fan shaped beam pattern is however complex in shape and in some cases mixture of fan and pencil shape beam pattern are also observed. The abrupt change in the correlation of the photon index with the flux also indicates the translation from super-critical regime to sub-critical regime \citep{Reig13}.

The recent outburst of the X-ray source 4U 1901+03 was detected on February 2019 by MAXI/GSC \citep{Nakajima19} and \textit{Swift}/BAT \citep{Kennea19}. \cite{Ji20} using \textit{Insight}-HMXT and \textit{NICER} observations found dozens of flares during 2019 outburst of the pulsar which were 1.5 times brighter than the persistent emission of the object. The shape of the pulse profiles during flares were found different from that of the persistent emission. However, at a comparable luminosities pulse profiles were similar to that of the persistent emission which indicates that the accretion onto the neutron star is only dependent on the mass accretion rate. \cite{Lei09} observed dependence of spectrum on phase and found that at beginning of the outburst the optical depth of Compton scattering was maximum near the major peak phase while during decay it was away from the main peak of the pulse profile. They observed that the flux of Fe emission line was independent of the phase suggesting the origin of this line to be accretion disk. Using torque models \cite{Tuo20} studied the correlation between intrinsic spin frequency derivative and bolometric flux. The authors also estimated the distance to the pulsar to be 12.4$\pm$0.2 kpc. \cite{Nabizadeh20} studied the spectral evolution of the source  using \textit{Insight}-HXMT and \textit{NuSTAR} observations. They also studied the 10 keV feature of the source and also observed 30 keV feature in the \textit{NuSTAR} data. \cite{Beri20} showed that this feature has fulfilled all the necessary condition for being CRSF.
\\
The motivation of the present paper is to investigate the Be X-ray source 4U 1901+03 using \textit{Swift}, \textit{NuSTAR} and \textit{NICER} observation during 2019 outburst. The bursts are different from the thermonuclear burst which is characterize by sharp rise and exponential decay. We estimated the average duration of flares and number of flares in each \textit{NuSTAR} observation and also studied the variation of pulse profile and pulse fraction with time and energy. We have simultaneously fitted \textit{NuSTAR} and \textit{Swift}-XRT spectra in a broad 0.5-79 keV range and also performed phase resolved spectral analysis to study the variation of spectral parameters with phase. We have study the variation of the 10 keV feature with luminosity and phase. The phase variation of 30 keV feature in last two \textit{NuSTAR} were studied.   

\section{Observation and Data reduction}

During the recent outburst \textit{Nuclear Spectroscopic Telescope Array} (\textit{NuSTAR}) observed the source four times. \textit{NuSTAR} consist of two co-aligned telescopes operating in the energy range  3-79 keV, each telescope has its own focal plane module consisting of a solid state CdZnTe detector \citep{Harrison13}. Observed data reduction were carried on \textsc{heasoft v6.26.1}. We use clean event files obtained from unfiltered event files making use of the mission specific command \textsc{nupipeline}. These cleaned event files were then used to extract light curves and spectra for analysis. We took a 90$^{\prime\prime}$ circular region around the bright region of the image as the source region  and consider another region of the same size away from the bright region as  background region. We use the above regions to extract light curves and the relevant spectra with the help of a tool \textsc{nuproducts}. The background subtraction of the light curves is done using \textsc{lcmath}, and  the light curves from two focal plane modules (FPMA \& FPMB) are combined using the same tool. The barycentric correction of the light curves are done using \textit{ftool} \textsc{barycorr}. All the spectra are fitted in \textsc{xspec}  \;v12.10.1f \citep{Arnaud96}. From hereafter the four \textit{NuSTAR} observations are referred as Obs1, Obs2, Obs3 and Obs4 respectively.  

The standard screening and reprocessing of \textit{Swift}-XRT unfiltered event files were done  using \textsc{xrtpipeline}. For our analysis, we took \textit{Swift}-XRT observations done in Windows Timing (WT) mode having good timing resolution. A circular region of 20 pixels around the optical position of 4U 1901+03 was considered as the source region, another region of the same area but away from the central region was taken as the background region. Using these region files we extracted light curves and spectra in \textsc{xselect}. An ancillary response (ARF) file was created using a \textsc{xrtmkarf}, whereas the response matrics file was obtained from the latest calibration database files. The background correction of the light curves was made with the help of \textsc{lcmath}. Finally using \textsc{xrtlccorr} we created light curves which is corrected from telescope vignetting and point spread function. After 58637.08 MJD \textit{Swift}-XRT observations were made in Photon Counting (PC) mode and the source was only detected in one observation. So we used \textit{NICER} observations made after 58637.08 to study the variation of pulse profile with time. The standard data screening and reduction were made using \textsc{nicerl2}. Light curves in 0.2-12 keV energy range and having a binning of 1 ms were extract from the \textit{NICER} clean event files. We also applied a barycentric correction to the \textit{Swift}-XRT and \textit{NICER} light curves.

\section{Analysis and Results}
\subsection{Light curves, pulse profiles and pulse fraction}

The four different \textit{NuSTAR} observations were taken during different stages of the outburst  are depicted in the Fig.1. Light curves having time resolution of 4 s were plotted using ftool \textsl{lcurve}. The flares in the light curves of the pulsar are observed in all observations (Fig.2). The duration of these flare were tens to hundreds of seconds. The bursts were more frequent and longer enough during the peak period of the outburst (\cite{Ji20},\cite{James11}). Considering a flare on to a part of light curve having count rate 3$\sigma$ level above the mean we estimated duration for first two observations for (5-6) per hour and for another two observations its was 3-4 per hour. The beginning and end of a flare is considered to be a lowest point of flare below the mean. The mean duration of burst for the first observation was $\sim$135 s whereas for the second observation the means was found $\sim$62.18 s. However, it is found that for the last two observations the mean burst duration were $\sim$98.13 and 95.35 s.
\begin{table*}
\centering
\begin{tabular}{c c c c c}
\hline
obsId &  Date of Obs (MJD) & Exposure (ks) & P$_{s}$\\
\hline
90501305001  & 58531.121 & 17.85 & 2.76415$\pm$0.00001\\
90502307002  & 58549.308 & 12.25 & 2.76152$\pm$0.00002\\
90502307004  & 58584.946 & 21.45 & 2.76211$\pm$0.00006\\
90501324002  & 58615.752 & 45.12 & 2.76054$\pm$0.00001\\
\hline
\end{tabular}
\caption{Table showing four \textit{NuSTAR} observations indicated by their observation IDs along with the date of observation, exposure and the pulse period of pulsar. The pulse period is obtained using light curves in 3-79 keV energy range.}
\end{table*}

\begin{figure}
\centering
\includegraphics[scale=0.3]{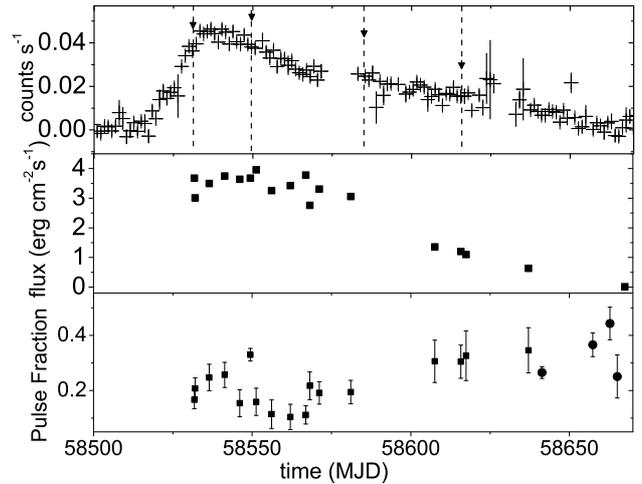}
\caption{Upper panel shows \textit{Swift}-BAT light curves of 4U 1901+03 during 2019 outburst in 15-50 keV energy range. The down arrows and vertical lines indicates four \textit{NuSTAR} observations. The middle panel shows variation of \textit{Swift}-XRT flux in 0.5-10 keV energy range. Fluxes are in the order of 10$^{-9}$ and obtained by using the command \textsf{flux} in XSPEC for \textit{Swift}-XRT spectra fitted by POWERLAW model. Bottom panel represents change in Pulse fraction with time, the square and circle symbols are for \textit{Swift}-XRT (0.5-10.0 keV) and \textit{NICER} (0.2-12 keV) respectively.}
\end{figure}
\begin{figure}
\centering
\includegraphics[width=8cm, height=9 cm]{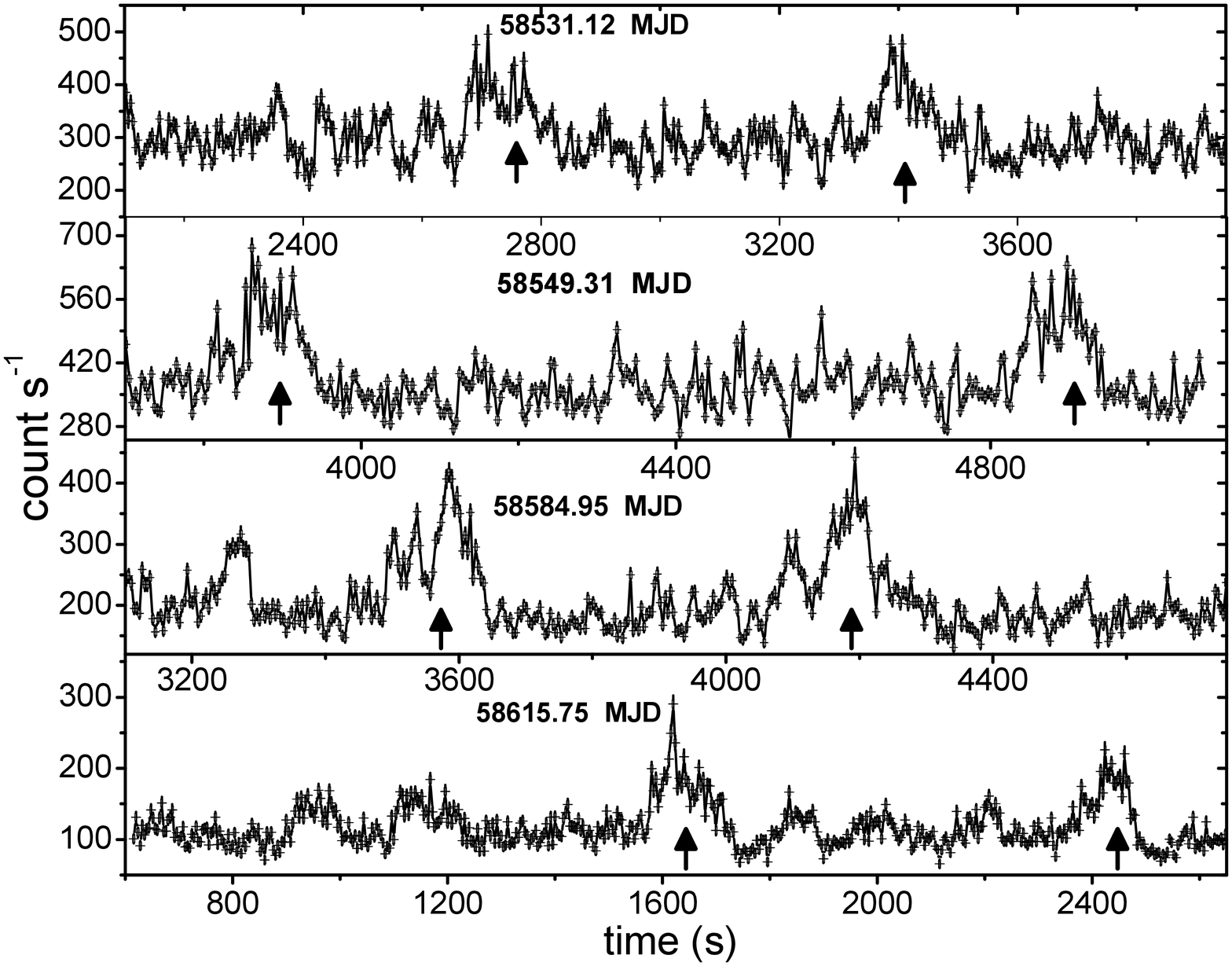}
\caption{Light curves showing some of the bursts observed in four \textit{NuSTAR} observations. The burst present in the light curves is indicated by up arrow.}
\end{figure}

 A crude estimation of the pulse period of the pulsar can be obtained by a Fourier transformation of the light curve. Therefore, we consider \textit{NuSTAR} light curves with binning of 1 ms to estimate the pulse period and to obtain the pulse profile. Final estimation of the pulse period was then obtained with the help of \textsc{efsearch} using the initial estimated value. The pulse periods for the four \textit{NuSTAR} observations is shown in table 1. The uncertainty associated with the pulse periods are estimated using the method described in \cite{Boldin13}. We can see from table 1 that the pulse period is not fixed but varies throughout the outburst. The evolution of the pulse period of an X-ray pulsar is caused by the transfer of the angular momentum from the accretion disk \citep{Ghosh79,Wang87}. During an outburst large amount of matters are accreted onto the neutron star along with the transfer of angular momentum which causes an accelerating torque to act on the neutron star. This accelerating torque causes an increase in the intrinsic spin frequency or decrease in the spin period of the pulsar. Pulse profiles are obtained by folding light curves in 3-79 keV energy range about the pulse period of the pulsar using the tool \textsc{efold}. For 0.5-10 keV \textit{Swift}-XRT we extracted light curves having binning of 0.0018 s and pulse profiles were extract using the above procedures. Pulse profiles were observed to be evolving with time. For \textit{NuSTAR} pulse profile at 58531.12 MJD is sinusoidal in shape (Fig.3 first column) which later evolved into double peaked with one main peak at 58549.31 MJD (Fig. 3 second column). At about 58584.95 MJD the pulse profile was consist of a single peak (Fig. 3 third column) having a notch and finally at 58615.75 MJD pulse profile becomes broad with single peak (Fig. 3 fourth column) but the shape is found different from that of the first observation (Fig. 3). The variation of the pulse profile with time were also studied using \textit{Swift}-XRT observations (Fig. 4). The pulse profile at 58532.03 MJD was two peaked with a strong primary peak. At 58541.33 MJD the second peak merged with the primary peak and at 58568.16 MJD the pulse profile was nearly single peaked with a notch. At 58581.05 MJD the pulse profile was a broad single peak. From 58607.54-58637.08 MJD pulse profiles were observed were almost the same and were having a broad single peak. The pulse profile at 58665.09 MJD is obtained by folding \textit{NICER} light curve and consist of a single peak which sharp compared to \textit{Swift}-XRT pulse profiles in between 58607.54-58637.08 MJD. After 58665.09 MJD no further pulsation was observed in the source. 
\begin{figure*}[t!]
\centering
\includegraphics[scale=0.5]{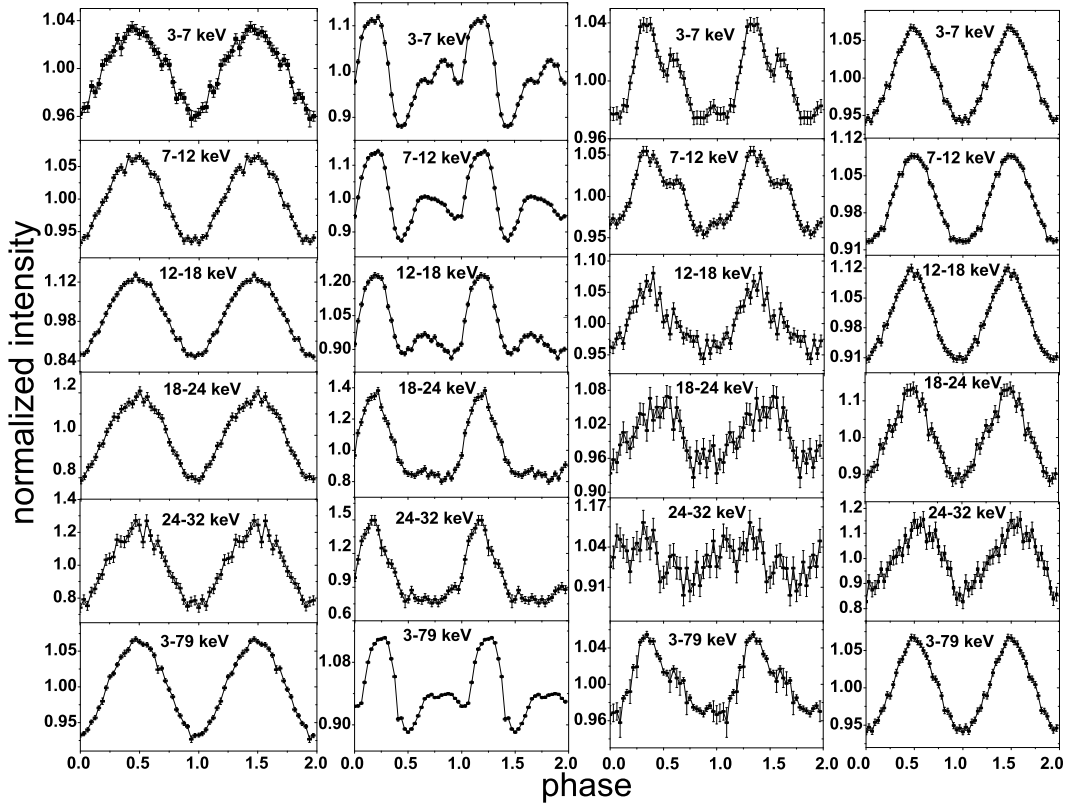}
\caption{Pulse profiles for different \textit{NuSTAR} observations and variation of pulse profile with energies. Figures (a),(b),(c) and (d) are for the 58531.12, 58549.31, 58584.95 and 58615.75 MJD respectively.}
\end{figure*}
\begin{figure}
\centering
\includegraphics[width=8 cm, height=10 cm]{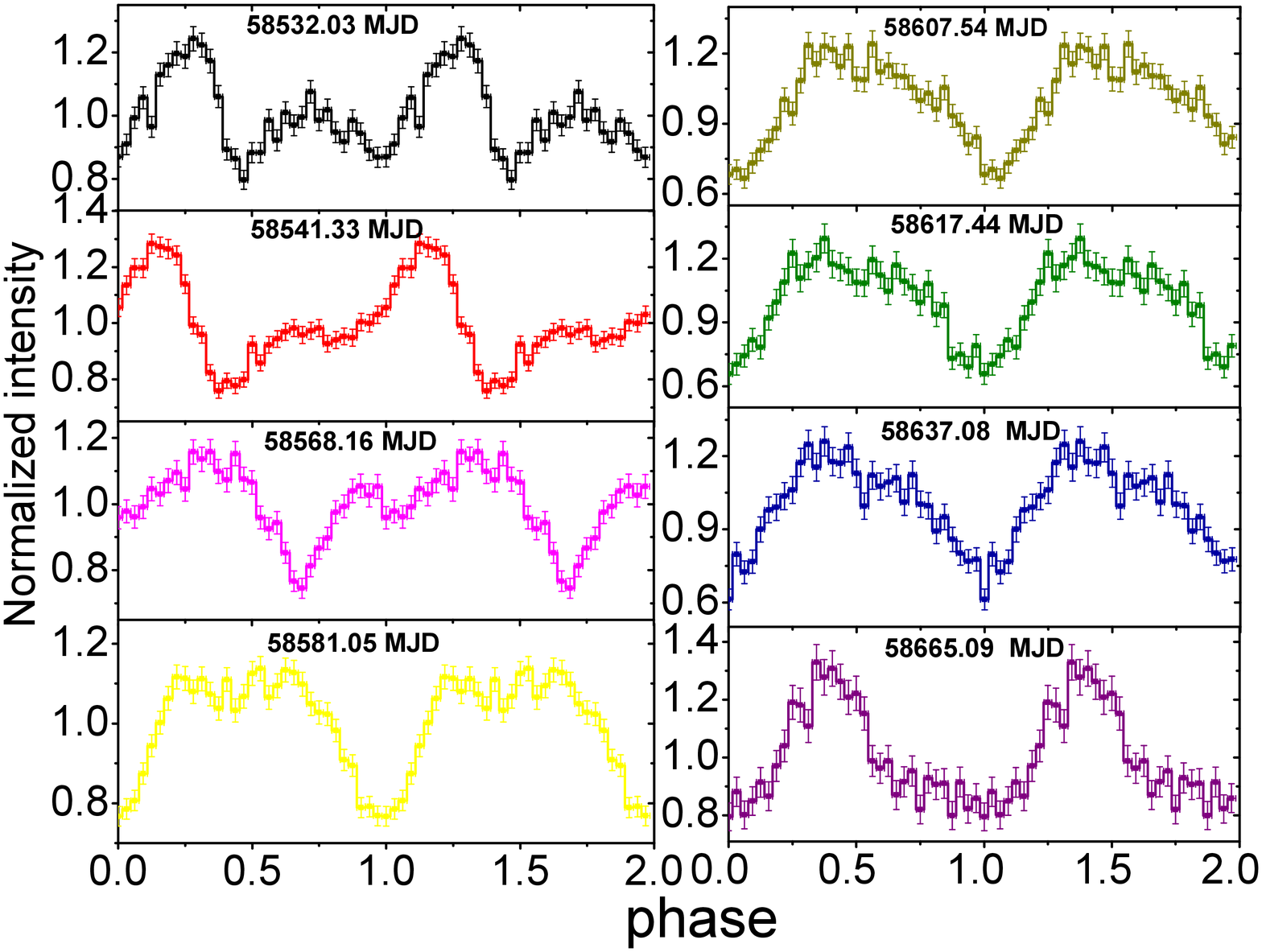}
\caption{Variation of Pulse profile with time. Pulse profiles from 58532.03 to 58637.08 MJD are obtained by folding \textit{Swift}-XRT lightcurve. For \textit{Swift}-XRT the significant level of pulse periods were greater than 3$\sigma$. The broadening of the \textit{\textit{Swift}}-XRT pulse profiles can be due to the short exposure time of the instrument so that there is not enough photons count rate to fold about the pulse period. The last pulse profile at 58665.09 MJD is obtained by folding \textit{NICER} lightcurve in 0.2-12 keV energy range (ObsId. 2200570141).}
\end{figure}
 In order to study the energy dependence of the pulse profile, we extract the light curves at different energy ranges using \textit{NuSTAR} observations. For Obs1 the pulse profiles at different energy bands are almost sinusoidal in shape (Fig. 3). Pulse profiles in the energy bands 7-12 and 3-79 keV are almost the same and as we move in hard X-ray region above 12 keV an increase in the height of the peaks of the pulse profiles is observed. For Obs2 (Fig. 3), pulse profiles in 3-7, 7-12, 12-18 and 3-79 keV are double peaked whereas none significant second peak was observed in the energy band 18-24 and 24-32 keV however an increase in height of the primary peak is observed in these energy bands. The 3-7 and 3-79 keV pulse profiles for Obs3 (Fig. 3) were having nearly a single peak with a notch near $\sim$0.5, above 7 keV the pulse profiles were not so smooth and were associated with large errors. In the case of Obs4 the pulse profiles at different energy bands were almost sinusoidal in shape, pulse profiles in energy bands 7-12 and 3-79 keV are almost the same however above 18 keV large errors were associated with each the of normalized count rates. 
 
  We define a pulse fraction (PF) parameter as $PF=(p_{max}-p_{min})/(p_{max}+p_{min})$,  where $p_{max}$ and $p_{min}$ are the maximum and minimum intensities of a pulse profile. In Fig. 5, PF is plotted with respect to energy for four \textit{NuSTAR} observations. The pulse fraction follows a complex variation with energy but there is an overall increase in its value with energy in all four observations. For Obs1 pulse fraction increases almost monotonically with energy (black). Pulse fractions in Obs2 increase slowly between 5-15 keV but above 15 keV pulse fraction increases rapidly (red). For the third observation (blue) pulse fractions increases slowly between 5-12 keV and remain almost constant between 12-21 keV and above 21 keV pulse fraction increases again. In the last observation there is overall increases in pulse fractions between 5-25 keV and above 25 keV the pulse fraction remained almost constant.  It is evident Fig. 5 that the pulse fraction is steeper for the first two observations when the luminosity was higher than the last two observations. Thus the pulse fraction in different energy ranges is high if the luminosity is high. The variation of the pulse fraction for different observations is also shown in table 2.

   The variation of the pulse fraction with time and flux were studied using \textit{Swift}-XRT observations. The pulse fraction was found to increase initially between 58531.77-58549.43 MJD from 0.17-0.33 and decreases to 0.16 at 58551.23 MJD and remain almost constant between 58556.14-58566.91 MJD. However in between 58568.16-58637.08 MJD the pulse fraction increases from 0.22 to 0.34 (bottom panel of fig 1). The variation of the pulse fraction with flux in 0.5-10 keV energy range is shown in bottom panel of figure 9. The pulse fraction decreases from 0.34-0.10 as flux increases from 0.64-3.43$\times$10$^{-9}$ erg cm$^{-2}$ s$^{-1}$ and increases from 0.11 to 0.33 between 3.43-3.68$\times$10$^{-9}$ erg cm$^{-2}$ s$^{-1}$ and abruptly decreases to 0.15 at 3.96$\times$10$^{-9}$ erg cm$^{-2}$ s$^{-1}$. Thus as we go below 3.43$\times$10$^{-9}$ erg cm$^{-2}$ s$^{-1}$ there is increase in pulse fractions. So as the outburst decays the pulse fraction increases.
\begin{figure}
\centering
\includegraphics[width=8 cm, height=6 cm]{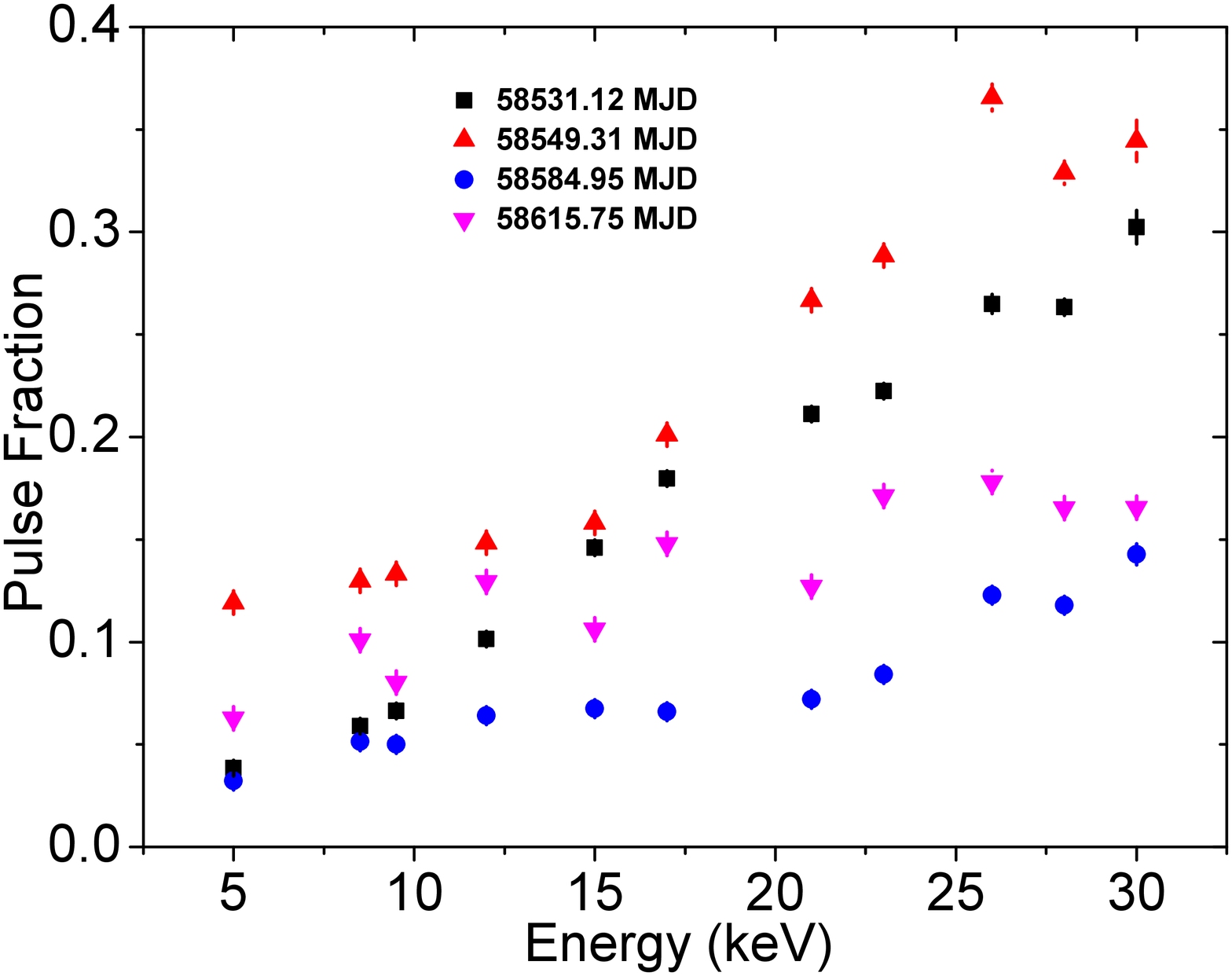}
\caption{Variation of the pulse fraction of the pulse profile with energy obtained using \textit{NuSTAR} observations.}
\end{figure}

\begin{table*}
\centering
\caption{Variation of pulse fraction with energy four different observations.}
\begin{tabular}{c c c c c}
\hline
Energy (keV) & Obs1 & Obs2 & Obs3 & Obs4 \\
\hline
5	& 3.85$\pm$0.15	 &	11.92$\pm$0.02	&	3.22$\pm$0.01 &	6.28$\pm$0.01\\
8.5	& 5.89$\pm$0.01	 &	12.98$\pm$0.34	&	5.13$\pm$0.01 &	10.08$\pm$0.03\\
9.5	& 6.63$\pm$	0.01 & 13.32$\pm$0.03	&	5.01$\pm$0.01 &	8.01$\pm$0.02\\
12	& 10.14$\pm$0.02 & 14.83$\pm$0.05	&	6.42$\pm$0.03 &	12.92$\pm$0.02\\
15	& 14.60$\pm$0.06 & 15.81$\pm$0.03	&	6.75$\pm$0.03 &	10.62$\pm$0.02\\
17	& 17.97$\pm$0.09 & 20.11$\pm$0.06	&	6.59$\pm$0.11 &	14.79$\pm$0.07\\
21	& 21.12$\pm$0.17 & 26.67$\pm$0.02   &   7.20$\pm$0.01 &	12.70$\pm$0.01\\
23	& 22.25$\pm$0.22 & 28.86$\pm$0.03	&	8.41$\pm$0.01 &	17.12$\pm$0.02\\
26	& 26.40$\pm$0.46 & 36.56$\pm$0.66	&	12.28$\pm$0.25 & 16.54$\pm$0.03\\
28	& 26.34$\pm$0.40 & 32.89$\pm$0.51	&	11.78$\pm$0.23 & 16.54$\pm$0.25\\
30	&30.24$\pm$0.82	 & 34.44$\pm$1.00	&	14.27$\pm$0.52 & 17.8$\pm$0.51\\

\hline 
\end{tabular}
\end{table*}

\subsection{Spectral Analysis}

\begin{table*}

\centering
 \begin{tabular}{c c c c c}
  \hline
Spectral parameters  & Obs1+\textit{Swift} & Obs2+\textit{Swift} & Obs3 & Obs4+\textit{Swift}\\
\hline
& & Model I & & \\
\hline
$n_{H}$ (cm$^{-2}$) & 3.80$\pm$0.12&4.03$\pm$0.07 &2.97$\pm$0.43 &3.74$\pm$0.23\\
$kT$ (keV) &0.20 $\pm$0.04 & 0.23$\pm$0.02& 0.31$\pm$0.05 &0.15$\pm$0.07\\
$\alpha$&0.51$\pm$0.02 &0.46$\pm$0.02 &0.34$\pm$0.01 & 0.80$\pm$0.02 \\
$E_{H}$ (keV)& 7.26$\pm$0.12&6.82$\pm$0.02 &6.21$\pm$0.08 & 6.87$\pm$0.02\\
$E_{Fe}$ (keV) & 6.52$\pm$0.03&6.52$\pm$0.07&6.49$\pm$0.02 &6.61$\pm$0.08\\
$\sigma_{Fe}$ (keV) &0.20$\pm$0.02&0.32$\pm$0.02 &0.30$\pm$0.07 &0.18$\pm$0.04\\
$E_{gabs_{1}}$ (keV)&10.71$\pm$0.11 &11.65$\pm$0.04 &10.14$\pm$0.01 &7.04$\pm$0.33\\
$\sigma_{gabs_{1}}$ (keV) &4.32$\pm$0.12 &2.49$\pm$0.09 &4.12$\pm$0.14 &6.54$\pm$0.23\\
$\tau_{gabs_{1}}$ &0.26$\pm$0.05& 0.18$\pm$0.03 & 0.33$\pm$0.05 &0.53$\pm$0.07\\
$E_{gabs_{2}}$ (keV) &... &... &30.37$\pm$0.55 &30.23$\pm$0.62\\
$\sigma_{gabs_{2}}$ (keV)&... &... &1.79$\pm$0.34 &1.04$\pm$0.35\\
$\tau_{gabs_{2}}$ &... &... &0.11$\pm$0.07&0.07$\pm$0.03\\
$flux$ (erg cm$^{-2}$ s$^{-1}$) &6.45$^{+1.21}_{-0.61}$&7.52$_{-0.74}^{+0.80}$ &5.01$_{-0.1}^{+0.23}$ &2.56$^{0.90}_{-0.41}$\\
$\chi_{\nu}^{2}$ & 1.02 & 1.11 & 0.98 & 1.01 \\ 
\hline
\hline 
& & Model II & &\\
\hline
$n_{H}$ (cm$^{-2}$) & 2.32$\pm$0.02&2.40$\pm$0.05 &2.70$\pm$0.13 &3.75$\pm$0.45\\
$kT$ (keV) &0.27 $\pm$0.05 & 0.39$\pm$0.04& 0.47$\pm$0.02 &0.28$\pm$0.08\\
$T_{0}$ (keV)& 1.27$\pm$0.07&1.08$\pm$0.08 & 1.33$\pm$0.17 & 0.75$\pm$0.11\\
$kT$ (keV) &4.81$\pm$0.05 &4.58$\pm$0.05 &4.52$\pm$0.06 & 4.63$\pm$0.11 \\
$\tau$&4.94$\pm$0.05 &4.89$\pm$0.13 &4.81$\pm$0.07 & 4.56$\pm$0.15 \\
$E_{Fe}$ (keV) & 6.52$\pm$0.07&6.45$\pm$0.02&6.61$\pm$0.07 &6.56$\pm$0.50\\
$\sigma_{Fe}$ (keV) &0.27$\pm$0.05&0.23$\pm$0.04 &0.24$\pm$0.04 &0.22$\pm$0.05\\
$E_{gabs_{1}}$ (keV)&11.32$\pm$0.36 &10.87$\pm$0.03 &9.57$\pm$0.64 &10.03$\pm$0.27\\
$\sigma_{gabs_{1}}$ (keV) &2.01$\pm$0.15 &1.57$\pm$0.21 &3.54$\pm$0.50 &3.68$\pm$0.12\\
$\tau_{gabs_{1}}$ &0.09$\pm$0.04& 0.07$\pm$0.02&0.17$\pm$0.05 &0.14$\pm$0.07\\
$E_{gabs_{2}}$ (keV) &... &... &30.09$\pm$0.57 &31.18$\pm$0.48\\
$\sigma_{gabs_{2}}$ (keV)&... &... &2.10$\pm$0.55 & 4.04$\pm$0.25\\
$\tau_{gabs_{2}}$ &... &... &0.15$\pm$0.07&0.10$\pm$0.06\\
$flux$ (erg cm$^{-2}$ s$^{-1}$) &6.41$^{+0.51}_{-0.54}$&7.49$_{-0.67}^{+0.61}$ &4.15$_{-0.70}^{+0.45}$ & 2.51$^{0.51}_{-0.25}$\\
$\chi_{\nu}^{2}$ &1.01 & 1.11 &1.00 &0.99\\ 
 \hline
\end{tabular}

\caption{Best fitted spectral parameters of 4U 1901+03 for four different cases using Model I and Model II. $\alpha$ and $E_{H}$ are photon index and highecut energy of the CUTOFFPL model. $E_{gabs}$ and $E_{Fe}$ are energy of absorption and Fe lines respectively. $\tau_{gabs}$ is the optical depth. $\sigma_{gabs}$ and $\sigma_{Fe}$ are the widths of absorption and Fe line. The subscript $gabs_{1}$ and $gabs_{2}$ are for two GABS models. The column density ($n_{H}$) and flux are in the scale of 10$^{22}$ cm$^{-2}$ and 10$^{-9}$ erg cm$^{-2}$ s$^{-1}$ respectively. $T_{0}$, $kT$ and $\tau$ are the spectral parameters of the COMPTT model. Flux were calculated in 3-79 keV energy range. Errors quoted are within 90$\%$ confidence interval.}
\end{table*}

The three \textit{NuSTAR} observations Obs1, Obs2 and Obs4 were close to three \textit{Swift} observations having obsIds 00088846001, 00088849001 and 00088870001 respectively. So we fitted \textit{Swift}-XRT and \textit{NuSTAR} (FPMA \& B) spectra simultaneously in 0.5-79.0 keV energy range, here \textit{Swift} spectra were in 0.5-10 keV energy range and \textit{NuSTAR} in 3-79 keV range.  However, there is a slight mismatch between the \textit{Swift}-XRT and \textit{NuSTAR} data points while fitting their spectra simultaneously which has been reported earlier by \cite{Bellm14}. A CONSTANT model was used while fitting XRT and \textit{NuSTAR} spectra simultaneously which take into accounts the instrumental uncertainties and also non-simultaneity of the observations.
The spectra were fitted with two different combinations of models. First, we used the combination of CONSTANT, PHABS, BLACKBODY, CUTOFFPL and GAUSSIAN and in the second case we replaced the CUTOFFPL with the model COMPTT which describes the Comptonization of the soft photon in hot plasma. Let us define Model I to be CONSTANT*PHABS*(CUTOFFPL+GAUSSIAN) and Model II to be CONSTANT*PHABS*(COMPTT+\\GAUSSIAN). The cross section for the PHABS was chosen to be \textsf{vern} and the abundance was set to \textsf{angr}. The optical depth of the Comtonizing region in Model II was obtained using disk geometry. However, in both cases, large negative residuals were observed near 10 keV. So we incorporated the Gaussian absorption model GABS in both cases. However, the HIGHECUT model did not fit the spectra well and large residuals were observed near the cutoff energy of the model.  

When Model I was used to fit Obs1 and Obs2 without GABS model a wave like a feature in the residuals between 3-30 keV energy range with large negative residuals near 10 keV were observed causing the fitting to be unacceptable (Fig. 6). An addition of GABS model fitted the spectra well. The reduced $\chi^{2}$ were about 1.76 and 1.52 for the first two cases respectively before the addition of GABS model which is unacceptable and after the addition it's values were about 1.02 and 1.11 respectively.

\begin{figure}
\includegraphics[scale=0.30,angle=-90]{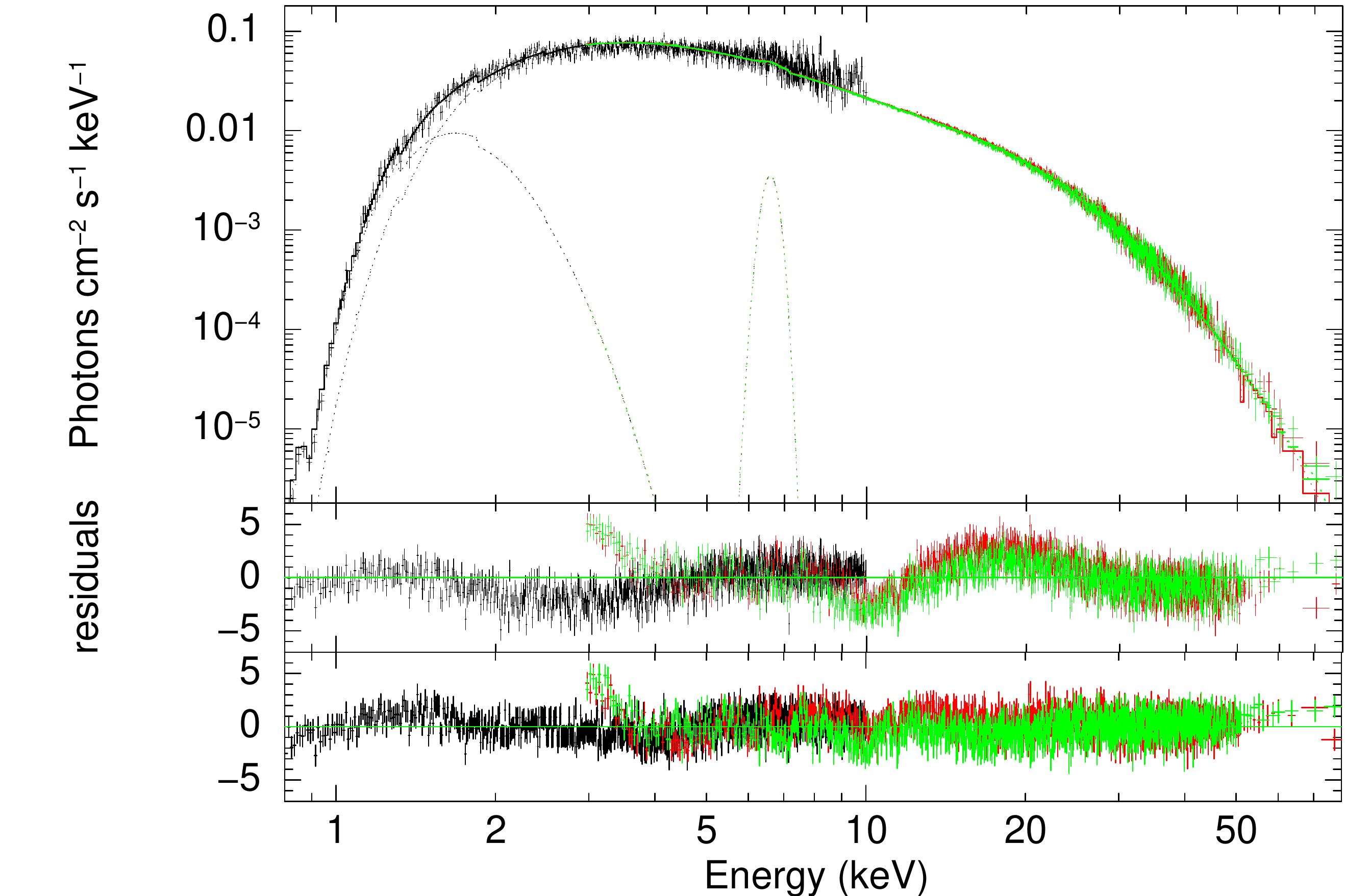}
\caption{Unfolded spectra of \textit{NuSTAR} Obs1 and \textit{Swift}-XRT observation fitted with Model I. The bottom and middle panels shows residuals after without and with the GABS model. Black indicates \textit{Swift}-XRT spectra, red and green indicates \textit{NuSTAR} FPMA \& B spectra respectively.}
\end{figure}

In the case of \textit{NuSTAR} Obs3 there were no \textit{Swift} observations close to the \textit{NuSTAR} observation, so we fitted 3-79 keV FPMA \& B spectra. The spectra were fitted well with Model I along with the GABS. As observed by \cite{Coley19} we too observed some negative residuals near 30 keV indicating another absorption like feature and possibly a Cyclotron Resonant Scattering Feature (CSRF). Fitting this absorption like feature with GABS model \cite{Coley19} found the energy of the line to be 31 keV with width 3.1 keV and optical depth about 1.1. So we added another GABS model and searched for an absorption feature near 30 keV (see Fig. 7), the best fitted value of line energy was 30.37 keV. The width and the depth of this absorption line were 1.79 keV and 0.11 respectively.

 The simultaneous fitting of \textit{Swift}-XRT and \textit{NuSTAR} Obs4 using Model I without GABS was not good as wave like feature was observed with large negative residuals near 10 keV. So we also used two GABS models one for 10 keV and another for 30 keV absorption like features, the spectra were fitted very well with reduced $\chi^{2}$ of the fitting about 1.01. However, the first absorption like feature was observed at 7.04 keV which is much below  what we have observed in three previous cases. The second feature was observed at 30.23 keV with width and optical depth about 1.04 keV and 0.07 respectively. 

\begin{figure}
\includegraphics[scale=0.30, angle=-90]{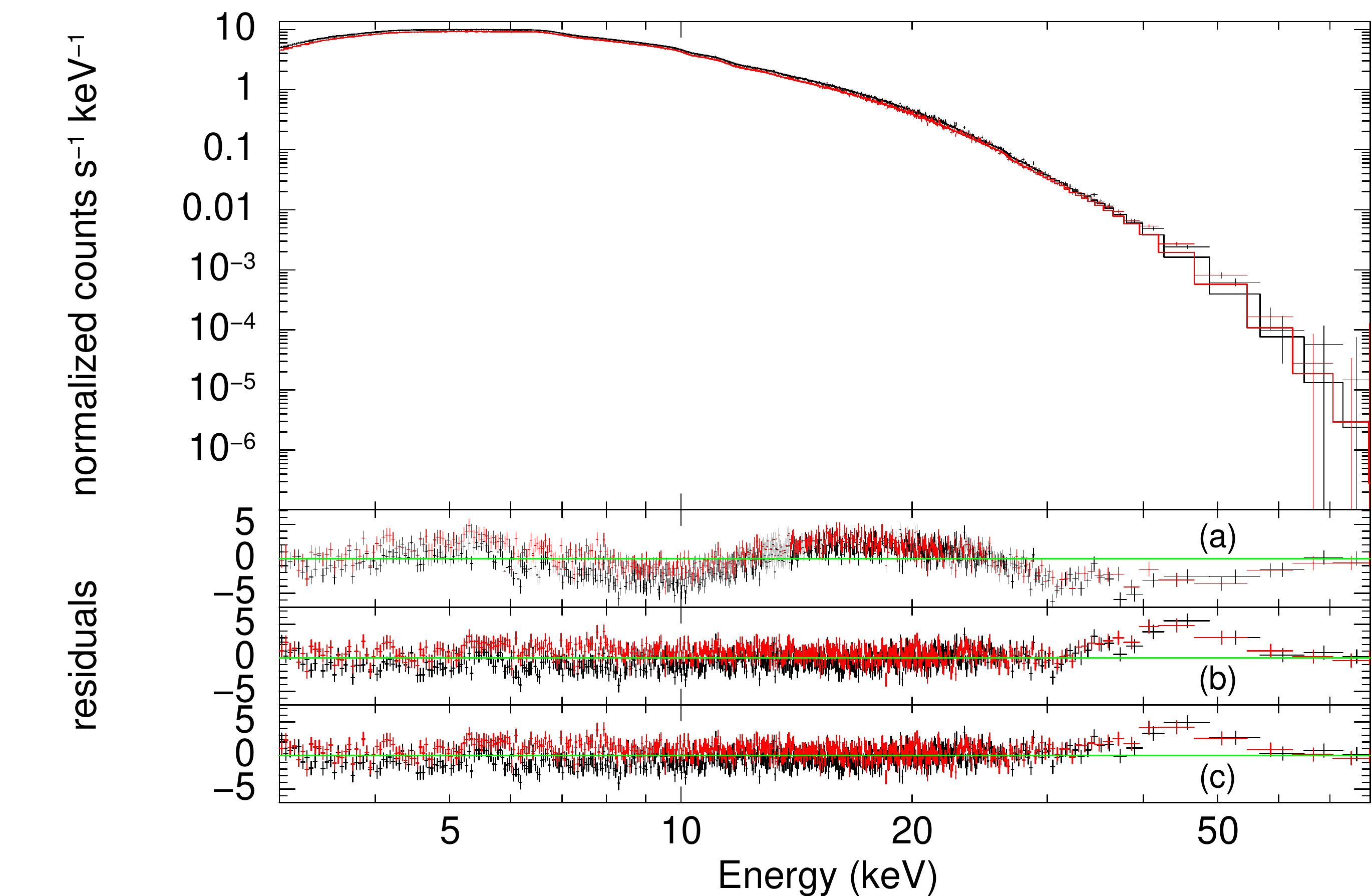}
\caption{Fitted spectra of \textit{NuSTAR} Obs3 in 3-79 keV energy range. Panel (a) shows residuals for Model I where as panel (b) is residuals for ModelI*GABS models and (c)shows residuals for ModelI*GABS*GABS models.}
\end{figure}

The spectra were also fitted well by Model II along with the GABS model. The column density is found to lie between 2.32-3.75$\times$10$^{22}$ cm$^{-2}$ (Table 3). The input soft temperature ($T_{0}$) of COMPTT model varies between 0.75-1.33 keV. The plasma temperature ($kT$) and the plasma optical depth ($\tau$) of this model are between 4.52-4.81 keV and 4.56-4.96 respectively. When the 10 keV feature was fitted with the absorption model the line energy was 11.32, 10.87, 9.57 and 10.03 keV for four \textit{NuSTAR} observations respectively. The width and the optical depth of this feature are in the range of 1.57-4.02 keV and 0.07-0.17. An absorption like feature around 30 keV was also observed in Obs3 and Obs4, so we used another GABS model and found the line energy of this feature to be about 30.09 and 31.18 keV.   

The spectral fitting by Model I estimated blackbody temperature was about 0.20-0.31 keV. The iron emission line was estimated to lie between 6.52-6.61 keV. For all four cases, the flux were estimated in 3-79 keV energy range. The best fitted spectral parameters are shown in table 3. The thermal component was also observed in the spectra fitted with Model II. The estimated flux were 6.45$\times$10$^{-9}$,7.52$\times$10$^{-9}$, 5.01$\times$10$^{-9}$ and 2.56$\times$10$^{-9}$ erg cm$^{-2}$ s$^{-1}$ respectively. Thus the luminosity of the pulsar in 3-79 keV were 1.24$\times$10$^{37}D^{2}_{4}$, 1.44$\times$10$^{37}D^{2}_{4}$,   9.59$\times$10$^{36}D^{2}_{4}$ and 4.89$\times$10$^{36}D^{2}_{4}$ erg s$^{-1}$ for four the cases respectively. The 10 keV absorption like feature was dependent on flux or luminosity and thus increases or decreases with the increase or decrease in flux or luminosity, from table 3 we can note that as flux decreases from 7.52$\times$10$^{-9}$ to 2.56$\times$10$^{-9}$ erg cm$^{-2}$ s$^{-1}$ the line energy of the feature decreases from 11.65 to 7.04 keV. Thus the line energy of the feature shows a positive correlation with the source luminosity. The absorption feature near the 30 keV was only observed in the last two \textit{NuSTAR} observations. This feature was also present in the spectra fitted with Model II. However, when the spectra were fitted with Model II the 10 keV feature does not show the same variation with flux or luminosity as seen in the case of spectral fitting with Model I but shows some positive correlation with flux or luminosity. The observed flux and $E_{Fe}$ were almost the same in the two cases. In order to check whether the origin of the 30 keV absorption like feature in spectra of the last two \textit{NuSTAR} observations was due to Compton reflection, we fitted  these spectra with the Compton reflection models like PEXRAV, PEXRIV and PEXMON. However, these models were not consistent with the observed spectra of the pulsar. So we rejected the possible origin of this feature due to Compton reflection.

\subsubsection{Fitting of \textit{Swift}-BAT spectra}
The 0.5-10 keV \textit{Swift}-BAT spectra were fitted with PHABS and POWERLAW models. The power law model fitted the spectra well and no additional models was required. As the pulsar slowly fades the photon index increase or in other words with decrease in flux the photon index increases (Fig. 9). Thus the spectra were softer near the end of the outburst. As the flux varies between 0.13-39.60$\times$10$^{-10}$ erg cm$^{-1}$ s$^{-1}$ the photon index varies between 1.03-2.1. The column density was observed to lie between 3.20-4.31$\times$10$^{22}$ cm$^{-2}$. The softening of the spectra of the pulsar at the end of an outburst was also observed by \cite{Reig16}.

\subsubsection{Phase Resolved Spectral Analysis}

In order to understand the variation of spectral parameters with pulse phase we performed phase resolved spectral analysis of the \textit{NuSTAR} observations. For phase resolved spectral analysis we have divided each pulse into 10 equal segments (Fig. 8). For each segment, a good time interval (gti) is created using \textsc{xselect} and using this gti file FPMA \& B spectra were produced. Each of the spectrum was fitted in 3-79 keV energy range with CONSTANT, PHABS, CUTOFFPL, GAUSSIAN and GABS models. Flux is estimated in the 3-79 keV energy range. Spectral parameters are found to vary significantly with the phases. From the phase-resolved spectroscopy of Obs1, the photon index ($\alpha$) and the highecut energy ($E_{H}$) were observed to show some anti-correlation with the flux. The flux varies between 6.40-7.02$\times$10$^{-9}$ erg cm$^{-2}$ s$^{-1}$ whereas $\alpha$ and $E_{H}$ varies between 0.19-0.366 and 6.08-6.64 keV respectively. The column density lies between 0.80-1.54 $\times$10$^{22}$ cm$^{-2}$. The variation of the Fe emission line ($E_{Fe}$) follows a complex pattern, its value decreases from 6.56 to 6.44 keV and then increase from 6.52 to 6.6 keV in between phase 0.2 to 0.6 and then decreases again, however, an abrupt increase in its value is observed between 0.9-1.0. The absorption like feature ($E_{gabs_{1}}$) was also observed to show some anti-correlation with the flux and lies between 10.12-11.02 keV. However the width ($\sigma_{gabs_{1}}$) and optical depth ($\tau_{gabs_{1}}$) of the line have two peak and also shows anti-correlation with the $E_{1gabs}$ and lies between 3.23-5.08 keV and 0.15-0.34 respectively. 

From phase-resolved spectral analysis of Obs2 the photon index and $E_{H}$ have some positive correlation with flux. Flux in this case decreases in between 0.0-0.3 from 7.89$\times$10$^{-9}$ to 7.79$\times$10$^{-9}$ erg cm$^{-2}$ s$^{-1}$, in between 0.4-0.8 the the flux increases reaching a maximum value 8.1$\times$10$^{-9}$ erg cm$^{-2}$ s$^{-1}$ at 0.7-0.8. The variation of the Fe line with the phase is complex. The absorption feature $E_{gabs_{1}}$ in between phase interval 0.0-0.1 is 11.38 keV and reaches a maximum value of 11.6 keV in between 0.1-0.2 and then decreases reaching a minimum of 11.07 keV at 0.9-1.0. $\sigma_{gabs_{1}}$ from 2.65 keV at 0.0-0.1 increases to reach a value of 3.12 keV after that it decreases reaching a minimum value of 2.06 keV in between phase 0.5-0.6 and increases then decreases again. The maximum value of $\sigma_{gabs_{1}}$ is 3.50 kev and was observed at 0.7-0.8. The optical depth ($\tau_{gabs_{1}}$) decreases from 0.15 at 0.0-0.1 to 0.10 at 0.6-0.7 and then increases abruptly to 0.17 and decreases again. The column density varies between 1.14-1.77$\times$10$^{22}$ cm$^{-2}$.

In Obs3 the variation of flux with phase is such that it exhibits two peaks one in between the phase 0.2-0.3 and another in between 0.7-0.8. The photon index and $E_{H}$ follows the flux and have two peaks. The Fe line from 6.58 keV decreases to reach a minimum value 6.43 keV in the phase interval 0.2-0.3 and increases reaching a peak in the interval 0.5-0.6 after that it again decreases and increases to reach a maximum value of 6.61 keV in the interval 0.9-1.0. From Fig. 8, we can see that the variation in $E_{gabs_{1}}$ is quite different from the two previous cases. The energy of the 10 keV feature $E_{gabs_{1}}$ decrease from 9.87 keV in between 0.0-0.3 followed by an increase and decrease. In the interval 0.4-0.6 the value $E_{gabs_{1}}$ increases again followed by a sharp decrease reaching 8.57 keV and after that, it increases again. As we can see from the figure the width of the line $\sigma_{gabs_{1}}$ shows negative correlation with the $E_{gabs_{1}}$. The optical depth $(\tau_{gabs_{1}})$ does not varies much in the phase interval 0.0-0.6, however its variation with phase is similar to that of the $E_{gabs_{1}}$ and varies between 0.28-0.46. In this case the column density varies between 0.38-1.72$\times$10$^{22}$ cm$^{-2}$.

From the phase-resolved spectral analysis of the fourth \textit{NuSTAR} observation, we found that the flux lies between 2.53-2.72$\times$10$^{-9}$ erg cm$^{-2}$ s$^{-1}$. The minimum value of photon index 0.39 was observed at phase interval 0.4-0.5 and the maximum value of 1.04 was observed at 0.5-0.6. The cutoff energy ($E_{H}$) was within 5.94-7.09 keV energy range. The value of column density ($n_{H}$) lies between 1.54-3.63$\times$10$^{22}$ cm$^{-1}$. The iron fluorescence line was lying between 6.45-6.64 keV with its width in 21-380 eV energy range. The 10 keV feature was found to vary between 5.84-8.79 keV. The width and the optical depth of the feature were lying between 1.97-5.59 keV and 0.68-0.99 respectively. 
 
The absorption like feature of about 30 keV in Obs3 was also found to depend on the pulse phase (Figure 10). The estimated line energy increases from 29.38 to 38.24 keV between the phase interval 0.0-0.4 and decreases to 33.38 keV, after that it increases to 33.88 keV and decreases again to 30.33 keV. In the phase interval 0.7-1.0 the line energy increases from 29.89-36.21 keV. The width $\sigma_{gabs_{2}}$ and optical depth $\tau_{gabs_{2}}$ of the feature also varies with the pulse phase and was found to lie between 2.17-7.16 keV and 0.2-0.7 respectively. For the Obs4 the 30 keV feature was found to vary between 27.6-33.51 keV with its width lying between 3.29-6.41 keV and depth varying between 0.17-0.42. 

We also fitted phase resolved spectra of Obs1 and Obs2 with GABS model in order to check the presence of 30 keV absorption like feature. For Obs1 the line energy and width of the feature varies between 31.24-38.76 keV and 0.75-6.15 keV respectively and in the phase interval 0.7-0.8 and 0.8-0.9 the feature was absent. In the case of Obs2 the energy of the feature is between 28.90-35.48 keV with a width lying between 0.77-6.06 keV. However for phase intervals 0.5-0.6 and 0.8-0.9 of Obs2 the value of $\sigma_{gabs_{2}}$ and $\tau_{gabs_{2}}$ were unrealistic so we did not considered there values.

\begin{figure*}[t!]
\centering
\includegraphics[scale=0.5]{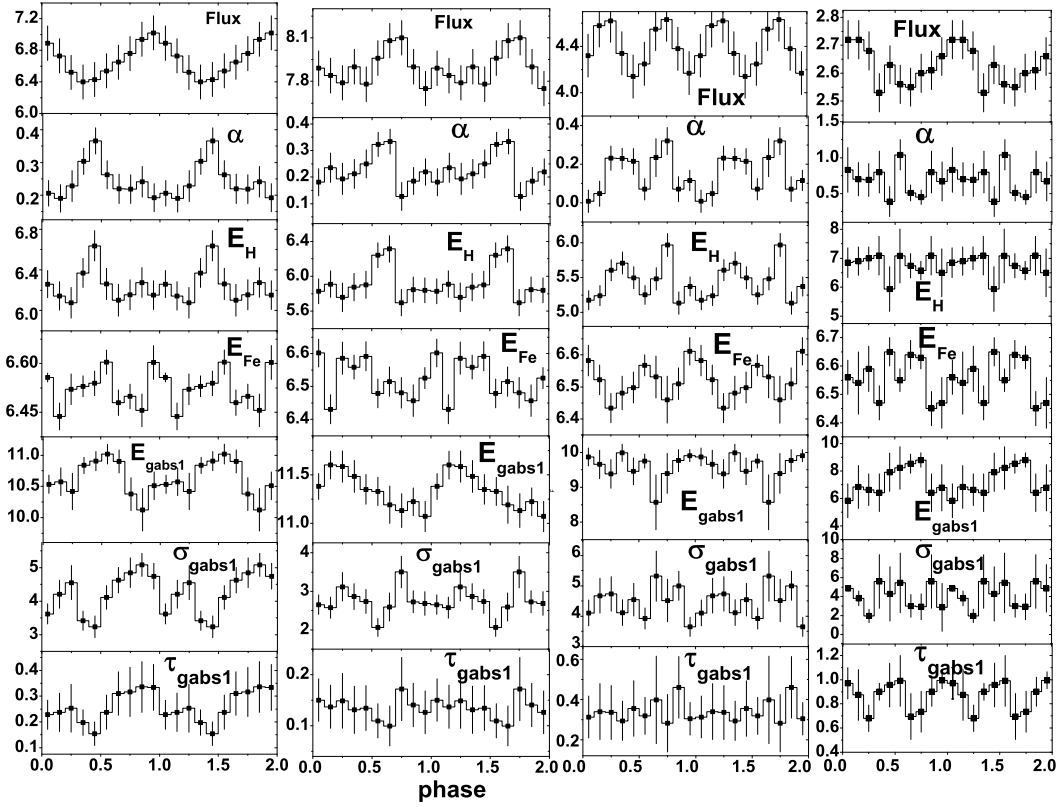}
\caption{Variation of Spectral parameters with phase for four \textit{NuSTAR} observations. Figures in first, second and third columns are for Obs1, Obs2, Obs3 and Obs4 respectively. $\alpha$ and $E_{H}$ are photon index and highecut of CUTOFFPL model. $E_{Fe}$ is the energy of Fe line. $E_{gabs_{1}}$, $\sigma_{gabs_{1}}$ and $\tau_{gabs_{1}}$ are the line energy, width and optical depth of GABS model used for fitting 10 keV feature. The error associated with the spectral parameter is standard error within 90$\%$ confidence interval.}
\end{figure*}

\begin{figure}
\includegraphics[scale=0.3]{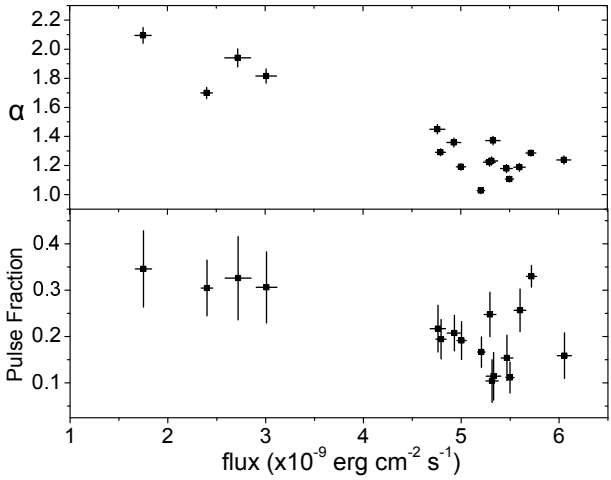}
\caption{Variation of photon index and pulse fraction with \textit{Swift}-XRT flux in 0.5-10 keV energy range.}
\end{figure} 

\begin{figure}
\centering 
\includegraphics[scale=0.3]{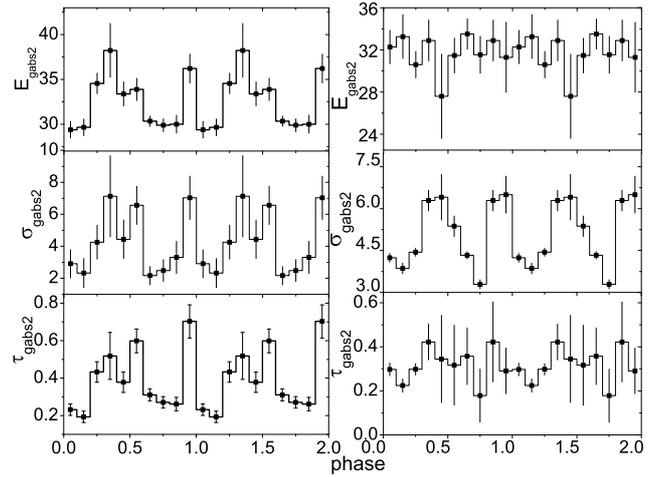}
\caption{Variation of 30 keV absorption feature with phase for Obs3 (first column) and Obs4 (second column).$E_{gabs2}$, $\sigma_{gabs2}$ and $\tau_{gabs2}$ are the line energy, width and the optical depth of the feature.}
\end{figure}

\section{Discussion}
We present Be X-ray pulsar 4U 1901+03 analysis which has undergone short bursts of tens to hundreds of seconds. The burst of the X-ray pulsar can be due to instability in accretion disk burst \citep{Taam88,Lasota91,Cannizzo96} similar to the burst observed in other sources namely, SMC X-1, GRO J1744-28 and MXB 1730-355 \citep{Moon03,Rai18,Fishman95,Lewin76}. Pulse profiles show variation with both time and flux and are similar to that observed by \citep{Lei09,Reig16,Ji20}. From the \textit{NuSTAR} observations, we found that the pulse profile having a single peak evolved into a double peak with one main peak and once again became a single peak (Fig. 3). Similar variations were observed in \textit{Swift}-XRT pulse profiles (Fig. 4). The height of the pulse profile peak increases with the increase in energy. Also, the double peaked pulse profile in Obs2 evolves into a single peaked at hard energy range. The pulse fraction is found to increase with the increase in energy and at the end of the outburst. As the X-ray emitting region gets smaller with an increase in energy and becomes more pulse as a result the pulse fraction increases \citep{Alexander08}. 

From the \textit{NuSTAR} observations, we observed that pulse profile (Fig. 3) is sinusoidal in shape at a luminosity of 1.24$\times$10$^{37}D^{2}_{4}$, which evolves into a double peak pulse profile with one main peak at 1.44$\times$10$^{37}$D$_{4}^{2}$ erg s$^{-1}$. The second peak of the pulse profile disappears and becomes a single-peak with a notch near $\sim$0.5 when the luminosity of the pulsar becomes 9.59$\times$10$^{36}$D$_{4} ^{2}$ erg s$^{-1}$. With the further decrease in luminosity, the notch disappears and the pulse profile becomes a single peak and this happens at 4.89$\times$10$^{36}$D$_{4}^{2}$ erg s$^{-1}$. Similar variations were observed in \textit{Swift}-XRT pulse profiles (Fig.4). Thus 4U 1901+03 shows luminosity dependent pulse profile having a double peak at high luminosity and a single peak at high luminosity. The complex variation of pulse profiles can be due to a change in emission beam pattern with luminosity (see section 1), which can be due to either fan and pencil beam pattern or mixture of these two beam pattern \citep{Chen08,Ji20,Reig16}. The fan shape beam pattern being dominated at the high luminosity whereas pencil beam pattern being dominating at the low luminosity states outburst \citep{Chen08,Reig16,Ji20}. The simple single peak pulse profile during Obs1 and Obs4 can be due  to the source in the sub-critical region where the pencil shape beam pattern dominates the fan shape beam pattern. The significant variation of pulse profile in Obs2 and Obs3 from  Obs1 and Obs4 can be due to the fan shape beam pattern or a mixture of fan and pencil shape beam pattern. We did not observe an abrupt change in the correlation between the flux and photon were seen suggesting the source was in which indicates that there was no transition between super-critical to sub-critical regimes. It might be possible that the source has not reached the pure super-critical regime during the outburst \citep{Chen08,Reig16}.  

The 10 keV absorption like feature was observed in all four \textit{NuSTAR} observations of the pulsar and was found to increase with the luminosity. Also, the width and optical depth of the feature varies for different observations. The energy, width and optical depth of the feature is within the range observed by \cite{Reig16}. It was observed that accreting pulsars show a positive correlation of the cyclotron line energy with the luminosity in sub-critical regime  and an egative correlation in super-critical regime \citep{Becker12,Mushtukov15}. If the pulsar was in sub-critical regime most of the time during the outburst as discussed above then the observed positive correlation of the 10 keV feature hints to be a cyclotron line. In addition to that the strong dependence of this feature on the viewing angle $i.e.$ on pulse phase like cyclotron line which also show a strong dependence on the pulse phase \citep{Isenberg98,Heindl04,Reig16} also support this feature to be a cyclotron line. The width and optical depth of the 10 keV features are within the range given by \cite{Coburn02} for other pulsars. The 10 keV feature was observed in pulsars having CRSF or not at all and was found to depend on the pulse phase \citep{Coburn02}. Considering the canonical value of neutron star parameters, the theoretically calculated value of critical luminosity was found by \cite{Becker12} to be $L_{crit}\sim1.49\times10^{37}B_{12}^{16/15}$, thus for this feature to be CRSF the critical luminosity must be $\sim10^{37}$ erg s$^{-1}$. Assuming the distance of the source to be 3 kpc \cite{Bailer18} observed luminosity lies between 2.69-8.04$\times$10$^{36}$ erg\;s$^{-1}$ which is below the critical luminosity. \cite{Reig16} noted that for the $L_{peak}/L_{crit}\sim1$ the distance should not be larger than $\sim$4 kpc. Thus for estimated luminosity to be less than the critical luminosity the distance of the source must be less than 4 kpc.  However, \cite{Strader19} noted that the distance of the object measured by the \cite{Bailer18} was not a well constraint because the parallax of the star in Gaia DR2 was insignificant and considering PS1 reddening maps along the direction of the source \citep{Green18} concluded that the distance must be greater than 12 kpc. Recently \cite{Tuo20} with the help of torque model and evolution of pulse profile during outburst estimated the distance of the source to be about 12.4 kpc. Assuming the distance of the source as 12.4 kpc the observed luminosity lies between 4.59-13.74$\times$10$^{37}$ erg s$^{-1}$ which is close to or above the critical luminosity and raises doubt about this feature being CRSF. \cite{Mushtukov15} showed that the critical luminosity is not a monotonic function of magnetic field and for pulsars having cyclotron energy about 10 keV the critical luminosity can reach a minimum value of few 10$^{36}$ erg s$^{-1}$. If this is the case then even if the source is at a distance of 3 kpc the observed luminosity will be at or above the critical luminosity. In \textit{NuSTAR} spectra weak residuals are observed around 10 keV due to tungsten L-edge of the \textit{NuSTAR} optics \citep{Madsen15,Furst13}. The 10 keV feature in other pulsars were present in the spectra of different instruments of different satellites (\cite{Coburn02}) like in the case of 4U 1901+03 where this feature was observed by RXTE \citep{Reig16} and \textit{Insight}-HMXT \citep{Nabizadeh20}, thus clearing doubt about the instrumental origin of the feature. We have also seen that the feature is present even if we used another continuum model COMPTT instead of CUTOFFPL.\cite{Nabizadeh20} showed that when \textit{NuSTAR} Obs3 spectra were fitted bya  two components model consisting of two POWERLAW*HIGHECUT along with GAUSSIAN and PHABS models no residuals were left near 10 keV and also no additional absorption model around 10 keV was needed when this two component model was used to fit \textit{Insight}-HMXT spectra. However, the authors also argued that transition from the typical cutoff power-law spectral shape to two-component spectral shape occurs at low luminosities about 10$^{34-36}$ erg s$^{-1}$, which indicates that the source distance must be small. Thus without proper estimation of the distance, we cannot be sure about the feature to be CSRF. It is equally possible that this feature  can be an inherent feature of the accreting X-rays pulsars or due to the departure of our phenomenological models used in fitting the spectra \cite{Coburn02}. The change in the hydrogen column density $n_{H}$ for different \textit{NuSTAR} observations or in phase-resolved spectral analysis can be as a result of absorption of photons by the interstellar medium.

As observed by \cite{Nabizadeh20} and \cite{Coley19} when Obs3 and Obs4+\textit{Swift} spectra were fitted some negative residuals were observed near 30 keV and fitting the spectra with absorption model we estimated the line energy about 30.37 and 30.23 keV for these observations respectively and were consistent with the value estimated by the authors. However, no negative residuals near 30 keV were observed in the first two \textit{NuSTAR} spectra which were having higher flux compared to the last two observations. In Obs3 and Obs4 the line energy of the 30 keV feature shows pulse phase variation. Despite the fact that this feature was not observed in the phase-averaged spectra of Obs1 and Obs2 it was observed in the phase-resolved spectra of these observations. However, in Obs1 and Obs2 this feature was not observable in some phases.  \cite{Beri20} have given sufficient evidence for this feature to be cyclotron line by studying the variation of line energy with luminosity and pulse phase. The authors also observed an abrupt change in the pulse profiles around the line energy of the feature. In X-ray pulsars with a high mass accretion rate, the accretion columns will appear to be confined by the strong magnetic field of the neutron star and are supported by internal radiation pressure. Thus observed cyclotron line can thus be originated from accretion column \citep{Schonherr14} or due to X-rays reflected from the neutron star’s atmosphere \citep{Poutanen13}. The absence of the cyclotron line in some observed energy spectra of the XRBs’ can possibly due to a large gradient of B-field strength over the visible column height or the latitude on the surface of a neutron star. The appearance of cyclotron line in certain pulse phases can be due to the partial eclipsing of the accretion column during which an observer is able to see some parts of the column \citep{Molkov19}. In such  a case, the magnetic field in the visible part of the accretion column is not so varied and we can observe cyclotron line in these phases like in the case of GRO J2058+42 \citep{Molkov19}. This can also be due to the gravitational bending of light as it affects the visibility of both the accretion columns and neutron star (see eg. \cite{Mushtukov18}).

When the magnetospheric radius $r_{m}$ becomes greater than the co-rotational radius $r_{co}$ then the centrifugal force will prevent the material from falling onto the neutron star this is known as Propeller Effect \citep{Illarionov75,Stella86}. As the propeller effect set in there is an abrupt decrease in the flux along with the absence of pulsation and even cause non-detection of the source. Here the co-rotational radius is defined as the radius where the keplerian angular velocity equals the spin angular velocity of the neutron star. The magnetospheric radius depends on the mass accretion rate, during a bright phase of an outburst the magnetospheric radius is less than the co-rotational radius so that matter can cross the magnetospheric radius and reach neutron star. As the mass accretion rate decreases the magnetospheric radius increase and can reach a point when this radius will become equal to the co-rotational radius and at this stage propeller phase sets in. From the \textit{NICER} observations, we found that the no pulsation was detected after 58665.09, also the flux abruptly decreases from 6.37$\times$10$^{-10}$ at 58637.08 MJD to 1.31$\times$10$^{-11}$ erg s$^{-1}$ at 58667.45 MJD which indicates that the pulsar has entered propeller phase. The increase in pulse fraction and the softening of the spectrum at the end of the outburst also support our argument \citep{Tsygankov16,Reig16,Zhang98}. As the accretion of matter onto the neutron star ceases when $r_{m}=r_{co}$ this implies that B=$4.8\times10^{10}P^{7/6}\left(\dfrac{flux}{10^{-9} erg s^{-1}}\right)^{1/2}\times\left(\dfrac{d}{1 kpc}\right)\left(\dfrac{M}{1.4 M_{\odot}}\right)^{1/3}\left(\dfrac{R}{10^{6} cm}\right)^{-5/2}$ G \citep{Cui97}), here $flux$ is the minimum bolometric X-ray when the pulsation was still detectable and $d$ is the distance to the source. Using \textit{Swift}-XRT flux 6.37$\times$10$^{-10}$ erg cm$^{-2}$ s$^{-1}$ in 0.5-10.0 keV observed at 58637.08 MJD, which is the minimum flux estimated in \textit{Swift}-XRT observations when the source was still pulsating and assuming the distance to the source to lie between 3-12.5 kpc the magnetic field of the neutron star lies 0.38-1.56$\times$10$^{12}$ G, assuming canonical values of mass and radius. However, the estimated magnetic field is associated with uncertainties as bolometric correction of the flux was not done and also the above minimum flux was not exactly known. Taking the cyclotron line energy to be 30 keV the estimated magnetic field of the neutron star will be about 2.59$\times$10$^{12}$ G and if it is so the distance to the source must be greater than 12.5 kpc.

 \section*{Acknowledgments}

The research has made use of \textit{NuSTAR}, \textit{Swift} and \textit{NICER} observational data and were obtained from the NASA High Energy Astrophysics Science Archive Research Center (HEASARC), Goddard Space Flight Center. The research is supported by SERB-DST research grant EMR/2016/005734. BR is thankful to DST and IUCAA Center for Astronomy Research and Development (ICARD), Physics Dept, NBU for extending research facilities. BCP is thankful to DST, New Delhi for  Project and IUCAA for Associateship program. 

\end{document}